\documentclass[preprint,aps,showpacs]{revtex4}
\usepackage{graphicx,amsfonts}
\usepackage{epsfig}
\usepackage{dcolumn}
\usepackage{bm}
\begin{document}

\title{Concerning the Landau Pole in 3-3-1 Models }
\author{Alex G. Dias\footnote{e-mail: alexdias@fma.if.usp.br}}
\affiliation{Instituto de F\'\i sica, Universidade de S\~ao Paulo, \\
C. P. 66.318, 05315-970 S\~ao Paulo, SP, Brazil. }

\author{R. Martinez\footnote{e-mail: remartinezm@unal.edu.co}}
\affiliation{Departamento de F\'\i sica, Universidad Nacional de
Colombia\\
Bogot\'a, Colombia. }

\author{V. Pleitez\footnote{e-mail: vicente@ift.unesp.br}}
\affiliation{Instituto de F\'\i sica, Te\'orica, Universidade Estadual
Paulista \\
Rua Pamplona 145, S\~ao Paulo, SP, Brazil. }
\date{\today}
\begin{abstract}
Some 3-3-1 models predict the existence of a non-perturbative regime at the TeV
scale. We study in these models, and their supersymmetric extensions, 
the energy at which the non-perturbative limit and a Landau-like
pole arise. An order of magnitude for the mass of the extra neutral vector
boson, $Z^\prime$, present in these models is also obtained.   
\end{abstract}
\maketitle

\section{Introduction}
\label{sec:intro}

The so called 3-3-1 extensions of the standard model (SM) are
interesting options for the physics at the TeV
scale~\cite{331,pt}. Although these models coincide at low
energies with the SM, they explain some fundamental questions that are
accommodated, but not explained by the former. For instance, i) in order
to cancel chiral anomalies the number of generation $N_g$ must be a multiple of
3, but because of the asymptotic freedom in QCD, which implies that the
number of generations must be $N_g\leq5$, it follows that in those
models the only number of generations allowed is $N_g=3$; ii) these models
as any with $SU(3)_W$ symmetry  explain why $\sin^2\theta_W<1/4$ at the
$Z$-pole (see below); iii) the electric charge is quantized independently of the
nature of neutrinos~\cite{pr}; iv) the Peccei-Quinn symmetry is almost an
automatic symmetry of the classical Lagrangian~\cite{pal} and with a minimal
modification, the PQ symmetry as the solution to the strong CP problem is
automatically implemented and the axion is protected against
semiclassical gravity effects~\cite{axion331}; v) the theory
becomes non-perturbative at the TeV scale, and the same happens with the
respective ${\cal N}=1$ supersymmetric version~\cite{331susy}. 
There are other models with $SU(3)_W$
symmetry~\cite{su3W}, but some of them imply charged heavy bileptons which are
stable, that in turn lead to potentially cosmological troubles~\cite{csaki}. 
Another interesting possibility to consider is by introducing extra
dimensions~\cite{extrad} and the orbifold
compactification~\cite{orbifoldsu3,orbifold331}, or other sort of 3-3-1 models
as in Refs.~\cite{outros331}. A common feature of models with $SU(3)_W$
electroweak symmetry is the existence of simply and doubly charged  or neutral
real or/and non-hermitian vector b\'osons. For instance, the doubly charged and the
real neutral vector bosons can be discovered by measurement of the left-right
asymmetries in lepton-lepton scattering~\cite{assi,tavares}, in
muonium-antimuonium transitions~\cite{willmann99,pleitez00} or in accelerator
processes~\cite{dion,maisbi}. 

One of the main feature of these models is the fact that
when the $g_L$ and $g_X$ coupling constants of the gauge groups $SU(3)_L$ and
$U(1)_X$, respectively, are related with the electroweak mixing angle, the
following relation is obtained
\begin{equation}
\frac{g^2_X}{g^{2}_L}=\frac{\sin^2\theta_W}{1-4\sin^2\theta_W}
\label{polo}
\end{equation}
in the models of Refs.~\cite{331,pt}. 
When $\sin^2\theta_W(\mu)=1/4$ the coupling constant $g_X(\mu)$ becomes
infinite, \textit{i.e.}, a Landau-like pole arises, however the theory loses
its perturbative character even at an energy scale lower than $\mu$. The other
possibility $g_L\to 0$ is ruled out since $g_L$ is the same as in the standard
model, $g_2$, due to the fact that the $SU(2)_L$ subgroup is totally embedded into
$SU(3)_L$.    

The possible existence of a Landau-like pole in 3-3-1 models is not
unexpected since every non-assintotically free theory seems to has such a
behavior. The new feature is that in some of these models that behavior may
happens at energies of just few TeVs. This will imply that the cut-off,
$\Lambda_{_{cutoff}}$, in the theory can not be eliminated, by taking
$\Lambda_{_{cutoff}} \to\infty$, as it is expected
in renormalizable theories. In this limit the theory might be a trivial theory.
This is supposed to be the case of pure QED since the works of Landau and
co-workers~\cite{landau}. From the phenomenological point of view this result is
not very dangerous, we already know that QED has to be embedded in the
electroweak theory at a few hundred GeVs, and also that weak and strong
correction have to be taken into account in the calculations of physical
observables, even those that are purely electromagnetic in origin, like the
$(g-2)_\mu$ factor, etc. 

However, as a mathematical laboratory it is interesting
to study pure QED at arbitrary small distances. In fact, lattice calculation
suggest that chiral symmetry breaking allows QED to escape the Landau pole
problem~\cite{qed1,qed2}. This should happen because the chiral
symmetry breaking is always strong enough to push the Landau pole above the
cutoff~\cite{qed1}. It is interesting that the possibility of the
e\-xis\-ten\-ce of the Landau pole, or that the theory is a trivial one, arises
already at the lower order in perturbation theory. More sophisticated
calculation only enhance our confidence that this phenomenon is not an effect of
the perturbation theory at lowest order. This is in accord with the point of
view that the renormalization group provide qualitative guidance with respect to
the asymptotic behavior at very high energies even where coupling constant at
the scale of interest are too large to allow the use of perturbation theory. In
particular this method provides usual insight into the types of possible
behavior in field theories~\cite{swbook}. 

Notwithstanding, we must remember that both
QED and the standard model are e\-ffec\-ti\-ve, and not fundamental, theories.
It means that effective operators with dimensions higher that $d=4$ have to be
considered if we want, for instance, to get a realistic continuum limit in
lattice calculations~\cite{swbook}. Thus it seem that using the pure versions of
these models are still inconclusive and the renormalization group may
give an insight in this issue in 3-3-1 models.

The possibility of triviality implies that new phenomena must enter
before the reach the Landau pole: or a new phase of the theory or the theory
must be embedded in a more general one. This has been used in the context of the
standard model to constraint the upper value of the Higgs boson
mass~\cite{hambye}. Moreover, recently it has been argued that this upper limit
on the Higgs bosons mass does not come from an instability of the
vacuum~\cite{holland}.  

In the past years some authors, using perturbation
theory, have calculated the energy scale at which the weak mixing angle get the
0.25 value~\cite{ng94,jain94,phf98}. They have been found, taking into account
only the degrees of freedom of the standard model, that this condition occurs at
an energy scale of the order of 3-4 TeV in the model of Ref.~\cite{331}. This
value is an upper limit of the energy scale at which the Landau-like pole
occurs (see the discussion in Sec.~\ref{sec:con}).    

Our goal in this paper is to study the running of $\sin^2\theta_W$ with
energy in 3-3-1 models~\cite{331,pt} and their supersymmetric
extensions given in Refs.~\cite{331susy}. However, since we have verified that,
with the representation content of the minimal models
$g_L$ does not change significatively, we study the running of $g_X$. 
We confirm the order of magnitude of the results of the previous works but
we considered a more general scenario, when the $SU(3)_L$ symmetry breakdown
scale, $\mu_{331}$, is inside the perturbative range and when the exotic quarks
or SUSY particles are considered much heavier than the other particles. 
As previous calculations, ours are also done at the 1-loop level but we briefly
comment the 2-loop case. For this reason our result have to be seen as
an estimative of the energy scale where
$\sin^2\theta_W=0.25$. Using of perturbation theory to find a
singularity could appears self-contradictory however, we recall that this
behavior at relatively low energy, arises because of the constraint in
Eq.~(\ref{polo}). We think that our calculations as the previous ones in 3-3-1
models are only preliminary results. When lattice calculations, or other more
appropriate techniques, were available
they could be compared with those 1-loop calculations, as is usually done
in the $\lambda\phi^4$ or QED cases~\cite{holland}. 

The outline of this paper is as follows. In Sec.~\ref{sec:sec2} we revise
briefly the minimal representation content of two 3-3-1 models that will be
considered next. In Sec.~\ref{sec:running} we give the evolution equation and
calculate the $b_i$ coefficients in each model with and without supersymmetry.
In this section we also reproduce the calculations of
Refs.~\cite{ng94,jain94,phf98} of the energy at which the condition $s^2_W=0.25$
is satisfied, taking into account only the degrees of freedom of the standard
model. In Sec.~\ref{sec:landau}, we study the evolution of $\alpha_X\equiv
g^2_X/4\pi$, and calculate the energy scale, $M^\prime$, at which
$\alpha_X(M^\prime)>1$. We also compare this energy with $\Lambda$ defined as
$\alpha_X(\Lambda)=\infty$, taking into account the degrees of freedom of the
3-3-1 models for energies above an energy scale that we denote $\mu_{331}$. The
last section is devoted to our conclusions. 

\section{3-3-1 Models with doubly charged vector bosons}
\label{sec:sec2}

The models that we will take into account in this section are characterized by
the electric charge operator, 
\begin{equation}
{\cal Q}_A=\frac{1}{2}\left( \lambda_3-\sqrt3\lambda_8\right)+X,
\label{q1}
\end{equation}
with two different representation content in the leptonic sector that 
are giving either by $\Psi_{aL}=(\nu_a,\,l_a,\,E^+_a)^T_L\sim({\bf1},{\bf3},0)$
(Model A) or by $\Psi_{aL}=(\nu_a,\,l_a,\,l^c_a)^T_L\sim({\bf1},{\bf3},0)$
(Model B), $\,a=e,\mu,\tau$. Both models contain doubly charged vector bosons. 
In the first case, we have to add singlets
$l_{aR}\sim({\bf1},{\bf1},-1)$, $E_{aR}\sim({\bf1},{\bf1},+1)$; and neutrinos
$\nu_{aR}\sim({\bf1},{\bf1},0)$, if necessary. In the second case 
only right-handed neutrinos have to be added, also if necessary. However,
since neutral singlet representations do not affect the running 
of the coupling constants, we will not worry about them here. In both
models the quarks transform as follows: 
$Q_{iL}=(d_i,\,u_i,j_i)^T_L\sim({\bf3},\,{\bf3}^*,
-1/3);\;i=1,2$; $Q_{3L}=(u_3,\,d_3,\,J)^T_L\sim({\bf3},{\bf3},2/3)$,
with the singlets $u_{\alpha R}\sim({\bf1},{\bf1},2/3)$, $d_{\alpha
R}\sim({\bf1},{\bf1},-1/3),\,\alpha=1,2,3$, $j_{iR}\sim({\bf1},{\bf1},-4/3)$,
and $J_R\sim({\bf1},{\bf1},5/3)$,

In these models there are fields with masses of the  order of
magnitude of the $SU(3)_L$ energy scale. For instance, in Model A the scalar
fields necessary to break the gauge symmetry down to 
$U(1)_Q$ and giving the correct mass to all fermions in the model are
three triplets:
$\eta=(\eta^0,\,\eta^{-}_1,\,\eta^+_2)^T\sim({\bf1},{\bf3},0)$,
$\rho=(\rho^+,\,\rho^0,\,\rho^{++})\sim({\bf1},{\bf3},1)$ and
$\chi=(\chi^-,\,\chi^{--},\,\chi^0)\sim({\bf1},{\bf3},-1)$.
We will denote the vacuum expectation values as follows:
$\langle\eta^0\rangle=u/\sqrt2$, $\langle\rho^0\rangle=v/\sqrt2$ and
$\langle\chi^0\rangle=w/\sqrt2$. In Model B it is necessary to add an scalar
sextet $S\sim({\bf1},{\bf6},0)$ 
\begin{equation} 
S=\left(\begin{array}{lll} 
\sigma^0_1 & h^-_1 & h^+_2\\ 
h^-_1 & H^{--}_1 & \sigma^0_2 \\ 
h^+_2 & \sigma^0_2 & H^{++}_2  
\end{array}\right), 
\label{sextet} 
\end{equation} 
and we will use the notation $\langle
\sigma^0_2\rangle=v_2/\sqrt2$. It is also possible to have $\langle
\sigma^0_1\rangle\not=0$ giving to the neutrinos a Majorana mass. We will not be
concerned with this here.  

In Model A the physical scalar spectra are such that a singly charged
scalar, which is a linear combinations of $\eta^+_1$ and $\rho^+$, has the
square mass~\cite{cp3} 
\begin{equation}
M^2_{1+}\propto \left(u^2+v^2-\frac{fwv}{u}-\frac{fuw}{v}\right),
\label{m1}
\end{equation}
and the other singly charged scalar is a linear combination of $\eta^+_2$ and
$\chi^+$ and has a square mass
\begin{equation}
M^2_{2+}\propto \left(w^2+v^2-\frac{fvw}{u}-\frac{fuv}{w}\right),
\label{m2}
\end{equation}
where $f<0$ is the trilinear coupling with dimension of mass. We see that even
if $f=0$ only one of the singly charged scalar has 
a low mass, the other one has a mass square proportional to $w^2$, so it is 
heavy enough (unless it is fine tuned) to be decoupled from low energy physics
(below the breaking of the $SU(3)_L$ symmetry). The same happens with the doubly
charged physical scalar, a linear combination of  $\rho^{++}$ and $\chi^{++}$,
with a mass square 
\begin{equation}
M^2_{++}\propto\left(w^2+v^2-\frac{fuv}{w}-\frac{fuw}{v}\right),
\label{m3}
\end{equation}
and we see that it is also too heavy (unless fine tuned) and it will not be
considered at low energies. On the other hand, in the real scalar fields sector
there are three physical scalar Higgs, two states that do not depend on $w$ and
$f$ having square masses of the order of $u^2+v^2$  and the other one is heavy.
It means that at low energies we consider two scalar doublets of $SU(2)_L$
when we use the standard model degrees of freedom. The vector bosons, $V^-$,
$U^{--}$, the exotic quarks and scalar singlets, all of them may be heavy
since their masses are proportional to $w$. The extra neutral vector boson
$Z_2$, has an even higher mass and also
it will not be considered at low energies. This vector bosons is a mixture of
$Z$ and $Z^\prime$, but if we neglect this mixture $Z_2\approx Z^\prime$. 

\section{The evolution equations}
\label{sec:running}

The evolution equations of the coupling constants at the one loop level are
given by 
\begin{equation}
\frac{1}{\alpha_i(M)}=\frac{1}{\alpha_i(M_Z)}+\frac{1}{2\pi}
\,b_i\,\ln\left(\frac{M_Z}{M}\right),\;i=1,2,3;
\label{rccgeral}
\end{equation}
where $\alpha_i=g^2_i/4\pi$ and $g_3,g_2,g_1$ are the coupling constant of the
$SU(3)_C,SU(2)_L,U(1)_Y$ 
gauge groups, respectively. In the context of 3-3-1 models
we define the $b_i$ coefficients corresponding to the coupling constants
$g_3,g_L,g_X$ of the gauge groups $SU(3)_C,SU(3)_L,U(1)_X$, respectively. 

For a general $SU(N)$ gauge group the $b_i$ coefficients are given by
\begin{equation}
b_i=\frac{2}{3}\sum_{\rm fermions}T_{Ri}(F)+\frac{1}{3}
\sum_{\rm scalars}T_{Ri}(S)-\frac{11}{3}\,C_{2i}(G)
\label{bi}
\end{equation}
for Weyl fermions and complex scalars, and $T_R(I)\delta^{ab}=\textrm{
Tr}[T^a(I)T^b(I)]$ with $I=F, S$; $T_R(I)=1/2$ for the fundamental
representation, $C_2(G)=N$ for $SU(N)$ and $C_2(G)=0$ for $U(1)$.
For $U(1)_y$ we use $\sum T_{R1}(F,S)=\sum y^2$ where
$y=Y/2$ for the standard model and $y=X$ for the 3-3-1 models. 
On the other hand for the respective supersymmetric version we have
\begin{equation}
b^{\textrm{susy}}_i=\sum_{\rm fermions}T_{Ri}(F)+
\sum_{\rm scalars}T_{Ri}(S)-3\,C_{2i}(G),
\label{bisusy}
\end{equation}
and only the usual non-supersymmetric fields are counted now.

We will assume that the standard model with several scalar multiplets is valid
until an energy scale $\mu_{331}$, \textit{i. e.}, below $\mu_{331}$ we consider
the SM plus some light scalar doublets or triplets. 
A $SU(2)_L\otimes U(1)_Y$ model with $N_g=3$ fermion generations, $N_H$ scalar
doublets ($Y=\pm1$) and $N_T$ non-hermitian scalar triplets ($Y=2$), using
Eqs.~(\ref{bi}) and the representation content above, implies 
\begin{eqnarray}
& &b_1=\frac{1}{6}N_H+N_T+\frac{20}{3},\nonumber \\
& & b_2=\frac{1}{6}N_H+\frac{2}{3}N_T-\frac{10}{3},\nonumber \\
& & b_3=-7.
\label{321}
\end{eqnarray}
Notice that we have not used a grand unification normalization for the
hypercharge $Y$ assignment. 
Since we will assume that $\mu_{\textrm{susy}}\approx \mu_{331}$ when
considering the SUSY extensions of a 3-3-1 model, below $\mu_{331}$ the only
effect of supersymmetry will be the addition of light scalar multiplets. Above
$\mu_{331}$, we have to consider the degrees of freedom of the 3-3-1 models. 

The heavy leptons, $E_a$ in Model A, quarks $J$ and
$j_i$; the scalar singlets $\eta^{+}_2$, $\rho^{++}$; the
scalar doublets like  $(\chi^{-},\chi^{--})$; and finally,
the vector doublets like $(V^-,Y^{--})$ and the extra neutral vector boson,
$Z^\prime$, will not be considered as active
degrees of freedom below $\mu_{331}$. Hence, in non-SUSY Model A we have
$N_H=2$ and $N_T=0$; in the SUSY version $N_H=4$ and $N_T=0$. 
In the non-SUSY Model B we have to take into account the scalar
sextet which implies an additional doublet and an non-hermitian triplet, so that
$N_H=3$ and $N_T=1$, and $N_H=6$ and $N_T=2$ in the SUSY case.  

\begin{figure}[ht] 
\begin{center} 
\leavevmode 
\mbox{\epsfig{file=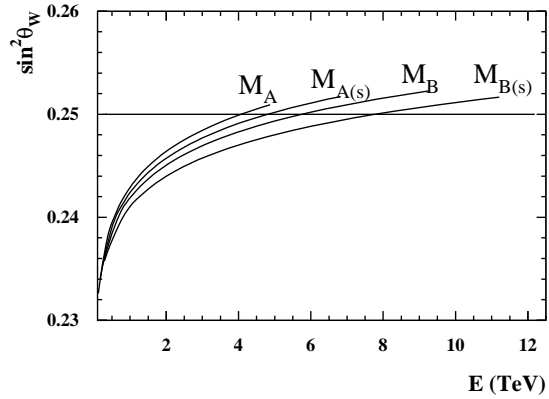,width=0.5
\textwidth,angle=0}}        
\end{center} 
\caption{Running of the electroweak mixing angle for Models A and B
considering only the degrees of freedom of the effective 3-2-1 model. 
(s) stands for the respective supersymmetrized version.}
\label{fig1} 
\end{figure}

In the energy regime below $\mu_{331}$ we use the standard definition for
$\sin^2\theta_W(\mu)$: 
\begin{equation}
\sin^2\theta_W(\mu)=\frac{1}{1+\frac{\alpha_2(\mu)}{\alpha_1(\mu)}},\quad
\mu \leq\mu_{331}.
\label{sin321}
\end{equation}
With this equation for the weak mixing angle and the $b_i$ coefficients in
Eqs.(\ref{321}), we obtain the value of
$\sin^2\theta_W(\mu_{331})$ using the running equation of $\alpha^{-1}_{1,2}$ 
given by Eqs.~(\ref{rccgeral}), which will be used as an input for the case of
energies above $\mu_{331}$. The values of the energies at which
$\sin^2\theta_W(\Lambda)=0.25$ (at which the curves cut the values
0.25), in models A and B with and without supersymmetry, are shown in
Fig.~\ref{fig1}. These energies give an order of magnitude of the energy scale
of the Landau Pole. In particular, for the non-SUSY Model A (leftmost curve in
the figure) we obtain a value for $\Lambda$ in agreement with that of
Refs.\cite{ng94,jain94,phf98}, i.e., $\approx 4$ TeV.

Next, we consider the case of energies above the scale $\mu_{331}$. Now we have
to use the relation  
\begin{equation}
\sin^2\theta_W(\mu)=\frac{1}{4}\,\frac{1}{1+\frac{\alpha_L(\mu)}
{4\alpha_{X}(\mu)}}\leq 1/4,\quad \mu \geq\mu_{331},
\label{sin331}
\end{equation}
and the running equation we will be concerned is 
\begin{widetext}
\begin{equation}
\frac{1}{\alpha_{X}(\mu)}= \left[1-4\sin^2\theta_W(M_Z) \right]
\,\frac{1}{\alpha(M_Z)}+\frac{1}{2\pi}\left(b_1-3b_2\right)
\,\ln\left( \frac{M_Z}{\mu_{331}}\right) +
\frac{1}{2\pi} {b^{r}}_{_X} \,\ln\left(\frac{\mu_{331}}{\mu}\right),
\label{alphax12}
\end{equation}
\end{widetext}
where $b^r_X$, with $r=A,B$, are given below by Eqs.~(\ref{bia}) 
for Model A, or by Eqs.~(\ref{bib}) for Model B, respectively; with
$\sin^2\theta_W(M_Z)=0.2311$, $\alpha(M_Z)=1/128$ and $M_Z=91.188$
GeV~\cite{pdg}. 

The $b_i$ coefficients  in Model A, with and without SUSY, when
the degrees of freedom above $\mu_{331}$ are taken into account, are given by:
\begin{eqnarray}
&& b^{A}_X= 24+N_\rho+N_\chi, \nonumber \\
&& b^{A}_{X\slash\!\!\!J}= 10+N_\rho+N_\chi, \nonumber \\
&& b^{A(\textrm{susy})}_X=36+3(N_\rho+N_\chi), \nonumber \\
&& b^{A(\textrm{susy})}_{X\slash\!\!\!J}= 15+3(N_\rho+N_\chi).
\label{bia}
\end{eqnarray}

Similarly, for the case of Model B, we have 
\begin{eqnarray}
&& b^{B}_X=20+N_\rho+N_\chi,\nonumber \\
&& b^{B}_{X\slash\!\!\!J}= 6+N_\rho+N_\chi,
\nonumber \\
&& b^{B({\textrm{susy}})}_X= 30+3(N_\rho+N_\chi),\nonumber \\
&&b^{B({\textrm{susy}})}_{X\slash\!\!\!J}=9+3(N_\rho
+N_\chi).
\label{bib}
\end{eqnarray}

In Eqs.~(\ref{bia}) and (\ref{bib}) we also show the cases when we omit the
exotic quarks (this is denoted by ${\slash\!\!\!J}$ in $b_X$). We recall that
in the supersymmetric versions we have assumed
$\mu_{\textrm{susy}}\approx\mu_{331}$. 
 
In the minimal non-SUSY version of models A and B, we have $N_\rho=N_\chi=1$ and
$N_\rho=N_\chi=2$ in the respective SUSY models. Notice that in both models
adding more triplet of scalars like $\rho$ and $\chi$ enhance $b_X$ and produce
a lower value for $\Lambda$. In the SUSY version the running is always faster.

\section{The Landau pole}
\label{sec:landau}

We see from Eq.~(\ref{sin331}) that $\alpha_{X}(\mu)\to \infty$ when
$\mu\to\Lambda$, and we have $\sin^2\theta_W(\mu)\to 0.25$. This is the Landau
pole that we have mentioned in Sec.~\ref{sec:intro}. In practice we study,
using Eq.~(\ref{alphax12}), what is the energy $\mu=M^\prime$ at which
$\alpha_{X}(M^\prime)>1$, \textit{i.e.}, the condition when $\alpha_X$ becomes
non-perturbative and we compare this result with the energy $\Lambda$ calculated
directly by the expression 
\begin{equation}
\Lambda=\mu_{331}\exp\left({\frac{2\pi}{b_X\alpha_X(\mu_{331})}}\right),
\label{pololandau}
\end{equation}
which must coincide with, or be of the same order, the value obtained by using
the condition $\sin^2\theta_W(\Lambda)=0.25$. Of course, we expect that
$M^\prime\lesssim \Lambda$.

\subsection{The Landau pole in Model A}
\label{subsec:lpma1}

As we said before, at energies below $\mu_{331}$ there is an
approximated $SU(3)_C\otimes SU(2)_L\otimes U(1)_Y$ symmetry with the 
particle content of the SM plus a second scalar doublet, \textit{i. e.}, in the
context of Model A: in the fermion sector there are the usual doublets
$(\nu_a,\,l_a)_L\sim({\bf1},{\bf2},-1)$ and $(u_\alpha,\,d_\alpha)_L
\sim({\bf3},{\bf2},1/3)$; the singlets $l_{aR}\sim({\bf1},{\bf1},-2)$ (and
right-handed neutrinos but they do not affect the running of the constants); and
$u_{\alpha R}\sim({\bf3},{\bf1},4/3)$, $d_{\alpha R}\sim({\bf3},{\bf1},-2/3)$.
In the scalar sector we have two doublets
$(\eta^0,\eta^-_1)\sim({\bf1},{\bf2},-1)$ and 
$(\rho^+,\rho^0)\sim({\bf1},{\bf2},+1)$.

These are the degrees of freedom that are active at energies 
$\mu\leq\mu_{331}$. Hence, we have
$N_H=2,N_T=0$ in the non-SUSY model and $N_H=4,N_T=0$ in the SUSY
one. With this particle content we have from Eqs.(\ref{321}) [for completeness
we include the coefficient $b_3$]: 
\begin{eqnarray}
&&(b_1,b_2,b_3)=(7,-3,-7), \nonumber \\
&&(b_1,b_2,b_3)^{\textrm{susy}}=(22/3,-8/3,-7), 
\label{num1}
\end{eqnarray} 
and using Eqs.~(\ref{sin321}) for obtaining
$\alpha_{X}(\mu_{331})$. If we use only the degrees of freedom that were used in
obtaining the coefficients above we get, using Eq.~(\ref{sin321}), that
$\sin^2\theta_W(\Lambda)=0.25$ when $\Lambda\approx4.10$ TeV in the
non-SUSY case and $\Lambda\approx4.8$ TeV in the SUSY case, as can be seen from
Fig.~\ref{fig1}. 

Above the $\mu_{331}$ scale, the full
representation of the 3-3-1 model have to be taken into account and we get,
according to Eqs.(\ref{bia}) [for future use we have included the coefficients
$b_L,b_3$],
\begin{eqnarray}
&& (b_X,b_L,b_3)^A=(26,-13/2,-5),\nonumber \\ 
&&(b_X,b_L,b_3)^A_{\slash\!\!\!\!J}=(12,-13/2,-7);\nonumber \\
&& (b_X,b_L,b_3)^{A(\textrm{susy})}=(48,0,0),\nonumber \\
&&(b_X,b_L,b_3)^{A(\textrm{susy})}_{\slash\!\!\!\!J}=(27,0,-3). 
\label{num2}
\end{eqnarray}

With the exotic quarks we will consider two situations. 
First, that they have masses below $\Lambda$ and are taken into account in
the evolution equations in the interval $[\mu_{331},\Lambda]$; second, we assume
that their masses are higher than $\Lambda$ and are not
considered in the running coupling constants.  

The result for the non-supersymmetric model appear in Table I for different
values for the $\mu_{331}$ scale: 2.0, 1.5, 1.0, 0.75 and 0.50 TeV. The same is
done in Table II for the supersymmetric model. In both cases the $M^\prime$
and $\Lambda$ values when the exotic quarks are considered heavy until the
Landau pole scale are shown in parenthesis in the respective table. In the last
column we show an order of magnitude of the $Z^\prime$ neutral vector boson (see
below).

\begin{table}
\begin{tabular}{||c|c|c|c|c||}\hline
$\mu_{331}$  & $\hat{\alpha}_X(\mu_{331})$ & $M^\prime$ & $\Lambda$
& $M_{Z^\prime}(\mu_{331}) $ \\ \hline
2.0 & 0.55 & 2.4(3.0) &3.1(5.2) & 5.2\\
1.5 & 0.39 & 2.2(3.4) & 2.8(5.7) & 3.3\\
1.0& 0.28 & 1.9(3.9) & 2.4(6.5)& 1.9\\
0.75  & 0.23 & 1.7(4.2) & 2.1(7.1)&1.3\\
0.5 &0.19 & 1.4(4.8) & 1.8 (8.2) &0.8\\ 
\hline
\end{tabular}
\caption{Values of $M^\prime$ and $\Lambda$ for the non-SUSY Model A. The
number inside parenthesis are the values for the case when we omit the exotic
quarks in the running equation. We show, in the last column, an estimative for
the mass of the $Z^\prime$ vector boson. All masses are in TeV.} 
\label{table1}
\end{table}

\begin{table}
\begin{tabular}{||c|c|c|c|c||}\hline
$\mu_{331}$  & $\hat{\alpha}_X(\mu_{331})$ & $M^\prime$  & $\Lambda$
 & $M_{Z^\prime}(\mu_{331})$ \\ 
\hline
2.0 & 0.47 & 2.3(2.6) & 2.6(3.3)& 4.9 \\
1.5 & 0.35 & 1.9(2.3) & 2.1(2.9) &3.1\\
1.0 & 0.26 & 1.4(1.9) & 1.6(2.4) &1.8\\
0.75  & 0.22 & 1.2(1.7) & 1.3(2.1)& 1.2\\
0.50 &0.18 & 0.9(1.4) & 1.0 (1.8) &0.75\\ \hline
\end{tabular}
\caption{Same as Table I but for the SUSY Model A.} 
\label{table2}
\end{table}

\subsection{The Landau pole in Model B}
\label{subsec:lpma2}

In this case below $\mu_{331}$
in the scalar sector we have to consider three doublets
$(\eta^0,\eta^-_1)\sim({\bf1},{\bf2},-1)$,
$(\rho^+,\rho^0),(h_2^+,\sigma_2^0)\sim({\bf1}, 
{\bf2},+1)$; and one non-hermitian triplet
$T\sim({\bf1},{\bf3},+2)$. 
With this particle content we have from Eqs.(\ref{321}), below $\mu_{331}$ 
\textit{i.e.}, we have $N_H=3$ and $N_T=1$ ($N_H=6,N_T=2$ in the SUSY
case):    
\begin{eqnarray}
&&(b_1,b_2,b_3)=(49/6,-13/6,-7), \nonumber \\
&&(b_1,b_2,b_3)^{\textrm{susy}}=(32/3,-1,-7).
\label{num3}
\end{eqnarray} 
If we use only the degrees of freedom that were used in
obtaining the coefficients above we get again, from Eq.~(\ref{sin321}), that
$\sin^2\theta_W(\Lambda)=0.25$ when $\Lambda\approx5.7$ TeV in the
non-SUSY case and $\Lambda\approx7.8$ TeV in the SUSY case, as can be  seen from
Fig.~\ref{fig1}. If we consider the
doublet $(h^+_2,\sigma^0_2)$ heavy (but keeping its VEV small) we get $b_1=8$ and 
$b^{\textrm{susy}}_1= 28/3$ and the values for $M^\prime$ and $\Lambda$
are a little bit smaller than the case considered here.

Above the $\mu_{331}$ scale, the full representation of the 3-3-1 model have to
be taken into account and we obtain, according to Eqs.(\ref{bib}) [again for future
use we have included again the coefficients $b_L,b_3$],
\begin{eqnarray}
&& (b_X,b_L,b_3)^B=(22,-17/3,-5),\nonumber \\
&&(b_X,b_L,b_3)^B_{\slash\!\!\!\!J}=(8,-17/3,-7),
\nonumber \\
&&(b_X,b_L,b_3)^{B(\textrm{susy})}=(42,5,0), \nonumber \\
&&(b_X,b_L,b_3)^{B(\textrm{susy})}_{\slash\!\!\!\!J}=(21,5,-3). 
\label{num4}
\end{eqnarray}
The results are shown in Table III for the non-SUSY Model B and in Table IV for
the respective SUSY model.

\begin{table}
\begin{tabular}{||c|c|c|c|c||}\hline
$\mu_{331}$ & $\hat{\alpha}_X(\mu_{331})$ & $M^\prime$  & $\Lambda$
&$M_{Z^\prime}(\mu_{331})$  \\ 
\hline
2.0 & 0.40 & 3.0(6.3) & 4.0(13.9)& 4.5\\
1.5 & 0.32 & 2.7(8.0) & 3.8(17.7) &3.0\\
1.0 & 0.24 & 2.4(11.3) & 3.2(24.8)&1.7\\
0.75  & 0.21 & 2.2(14.4) & 2.9(31.5)&1.2\\
0.5 &0.17 & 1.9(20.1) & 2.5 (44.2) &0.7\\ \hline
\end{tabular}
\caption{Same as Table I but for the Model B.} 
\label{table3}
\end{table}

\begin{table}
\begin{tabular}{||c|c|c|c|c||}\hline
$\mu_{331}$  & $\hat{\alpha}_X(\mu_{331})$ & $M^\prime$  & $\Lambda$
& $M_{Z^\prime}(\mu_{331})$  \\ 
\hline
2.0 & 0.34 & 2.7(3.5) & 3.1(4.8)&4.1\\
1.5 & 0.28 & 2.2(3.2) & 2.5(4.4)&2.8\\
1.0 & 0.22 & 1.7(2.8) & 1.9(3.8)&1.7\\
0.75  & 0.19 & 1.4(2.5) & 1.6(3.4)&1.1\\
0.5 & 0.17 & 1.0(2.2) & 1.2(3.0)&0.7 \\ \hline
\end{tabular}
\caption{Same as Table II, but for the SUSY Model B.} 
\label{table4}
\end{table}

\section{$Z^\prime$ mass and 2-loop evolution equations}
\label{sec:zp2}

The $Z^\prime$ is the heaviest vector boson of the models, the Landau pole
energy scale ($\Lambda$) is supposed to be an upper limit for its mass in the
context of a perturbative approach~\cite{ng94}.
However, the mass of this boson, at the scale $\mu_{331}$, and assuming
$\langle\chi^0\rangle\approx\mu_{331}$, has an order of
magnitude given by 
\begin{equation}
M_{Z^\prime}(\mu_{331})\simeq [4\pi\,\alpha_{X}(\mu_{331})]^{1/2}\,\mu_{331}.
\label{zprime}
\end{equation}
The values for the estimative of $M_{Z^\prime}$ using
Eq.~(\ref{zprime}) are shown in the last column of Tables
\ref{table1}--\ref{table4}. We see that for some values of $\alpha_{X}$,
$M_{Z^\prime}$ is larger than $M^\prime$ or $\Lambda$.

It is interesting to note that 
in the SUSY  version of Model A at energies above $\mu_{331}$ the dependence
with the energy in $SU(3)_L$ and $SU(3)_C$ is lost since, as can be seen from
Eq.~(\ref{num2}), at the 1-loop we have that $b_L=b_3=0$, \textit{i.e.},  
\begin{equation}
\alpha_L(\mu>\mu_{331})=\alpha_3(\mu>\mu_{331})=\textrm{constant},
\label{af1}
\end{equation}
and the same occurs for SUSY Model B for $\alpha_3$ as shown in
Eq.~(\ref{num4}). We can wonder if this is an artifact of the 1-loop
approximation. Thus, let us consider the 2-loop evolution equations that are
given by 

\begin{equation}
\mu\frac{d\, \alpha_i(\mu)}{d\, \mu}= \frac{1}{2\pi}  
\left [ b_i+\frac{1}{4\pi}\sum_{j=1}^{3} b_{ij}
\alpha_j(\mu)\right ]\alpha_i(\mu)^2,
\label{rg2loop}
\end{equation}
where we have not considered the Yukawa couplings since in any case their 
dominant contributions seems to be positive. For example, in
SUSY Model A~\cite{jones} 
\begin{eqnarray}
b_{ij}= \left(\begin{array}{ccc}
\frac{784}{3}+12(N_\rho+N_\chi)& 32+16(N_\rho+N_\chi)& 160 \\
4+ 2(N_\rho+N_\chi) & 14+\frac{17}{3}(N_\rho+N_\chi) & 48\\
  \frac{62}{3} & 24 & 48
\end{array}\right).
\label{bij}
\end{eqnarray}
In fact, in Eqs.~(\ref{rg2loop}), we have $N_\rho=N_\chi=2$ since we are 
considering the SUSY version.
We see that even at this order the asymptotic freedom for QCD has been lost for
energies higher than $\mu_{331}$. The lose of the asymptotic freedom at higher
energies, for both the SUSY Model A and B is a prediction of the models
since this result does not depend on the value of $\mu_{331}$. Of course, a
more careful analysis should be done.

\section{Discussions}
\label{sec:con}

We have re-examined the question of the non-perturbative limit and the
Landau-like pole in 3-3-1 models. 
In addition we have considered  the respective supersymmetric versions and also
the situation when the exotic quarks are heavy enough and 
do not enter in the running equation of $\alpha_{X}$.
In practice what we have studied is the energy scale at which a model loses its
perturbative character, $M^\prime$, or calculated directly
the Landau pole, $\Lambda$, from Eq.~(\ref{pololandau}). 
We find, as expected, that for all these models these energy scales are of the
same order of magnitude, \textit{i.e.}, $M^\prime\lesssim \Lambda$. 

From Table~\ref{table1} we see that for Model A, the values of $M^\prime$ and
$\Lambda$ decrease with the value of $\mu_{331}$ but increase for lower
$\mu_{331}$ if we omit the exotic quarks in the running equation. The
maximal values of 4.8 TeV or 8.2 TeV without the exotic quarks, respectively,
occur when $\mu_{331}=500$ GeV. For the respective SUSY cases, we see from
Table~\ref{table2} that $M^\prime$ and $\Lambda$ always decrease with
$\mu_{331}$ and also that they have lower values than the respective non-SUSY
model. The result for the model with the scalar sextet (Model B) are shown in
Tables~\ref{table3} and \ref{table4}. The largest value for $M^\prime$
($\Lambda$) is 20.1 (44.2) TeV when the heavy quarks are not considered. As in
Model A, both scales also decrease with the value of $\mu_{331}$. 

Notice that from Table~\ref{table1}, the value of $\Lambda$ (or
$M^\prime$) for Model A (without SUSY) is always lower that the value obtained
in Refs.~\cite{ng94,jain94,phf98} and in Fig.~\ref{fig1}. As we have mentioned
before, the latter value should be an upper limit for $\Lambda$. This is
confirmed when the extra degrees of freedom of the 3-3-1 model are taken into
account, for energies above $\mu_{331}$. As said before, we can  see from
Tables~\ref{table1}--\ref{table4}, that the value of $\Lambda$ increases when
the scale $\mu_{331}$ increases. But, as $\mu_{331}$ becomes 
larger the difference between both energy scales becomes smaller and in some
point $\mu_{331}$ must be equal to $\Lambda$. However, notice also that when we
omit from the analyses the exotic quarks this upper bound is evaded. This
happens because the right-handed components of those quarks have the largest
value of the $U(1)_X$ charge making that $\alpha_X$ run more rapidly compared to
the case where only the standard model particles are taken into account. When
the heavy quarks are switch off $\alpha_X$ run again slowly and the Landau-like
pole occurs at a higher energy.
The scenarios without the exotic quarks could be realized if there are
strong dynamical effects with these degrees of freedom in this range of energy
an probably the number of scalar multiplets of these models may be lower
than it has been considered~\cite{dasjain}. 
Notice also that since SUSY implies
more degrees of freedom the values of $M^\prime$ and $\Lambda$ are always lower
than in the respective non-SUSY model. If the Landau pole is calculated by using
only the degrees of freedom below the 3-3-1 energy and the condition
$\sin^2\theta_W(\Lambda)=0.25$ from Eq.~(\ref{sin321}), the value obtained is
shown in Fig.~\ref{fig1}.  Since the value of $\mu_{331}$ is below of these
values we have studied how the value of the pole and the perturbative limit are
modified when $0.5\leq \mu_{331}\leq 2$ TeV.

Finally, let us mention that there is another type of 3-3-1 model in which the
right-handed neutrinos or heavy neutral leptons belong to
the same triplet than the ordinary leptons~\cite{mpp,tavares2}. The charge
operator is defined in this case as 
\begin{equation}
{\cal Q}_B=\frac{1}{2}\left(
\lambda_3-\frac{1}{\sqrt3}\lambda_8\right)+X.
\label{q2}
\end{equation}
In this sort of models the Landau pole arise above the Planck scale
and for this reason it has no physical consequences.
However, models with electric charge operator defined by
Eqs.~(\ref{q1}) and (\ref{q2}) are embedded in an $SU(3)_C\otimes SU(4)_L\otimes
U(1)_N$ but in this 3-4-1 model the equation relating the coupling constant
$g_L$ and $g_X$ is given also by Eq.~(\ref{polo})~\cite{su4}. Thus, our results
are also valid for the case of 3-4-1 models.

A. G. D. was supported by FAPESP under the process 01/13607-3, R.M. was
supported by COLCIENCIAS and V. P. was partially supported by CNPq under the
process 306087/88-0.


\begin{thebibliography}{99}

\bibitem{331} F. Pisano and V. Pleitez, Phys. Rev. D \textbf{46} (1992) 410;
R. Foot, O. F. Hernandez, F. Pisano and V. Pleitez, Phys. Rev. D \textbf{47}
(1993) 4158;
P. \ Frampton, Phys.\ Rev.\ Lett.\ \textbf{69} (1992) 2889.
\bibitem{pt} V. Pleitez and M. D. Tonasse, Phys. Rev. D \textbf{48}, 2353
(1993).
\bibitem{pr} C. A. de S. Pires and O. Ravinez, Phys. Rev. D \textbf{58}, 035008
(1998). 
\bibitem{pal} P. B. Pal, Phys. Rev. D \textbf{52}, 1659 (1995).
\bibitem{axion331} A. G. Dias, V. Pleitez, and M. D. Tonasse, Phys. Rev. D
\textbf{67}, 095008 (2003); A. G. Dias and V. Pleitez, Phys. Rev. D \textbf{69},
077702 (2004); A. G. Dias, C. A. de S. Pires, and P. S. Rodrigues
da Silva, Phys. Rev. D \textbf{68}, 115009 (2003) and references therein.
\bibitem{331susy} J. C. Montero, V. Pleitez and M. C. Rodriguez,
Phys. Rev. D \textbf{65}, 035006 (2002); \textit{ibid} D\textbf{70}, 075004
(2004), and hep-ph/0406299.
\bibitem{su3W} S. Dimopoulos and D. E. Kaplan, Phys. Lett. \textbf{B531}, 127
(2002).
\bibitem{csaki} C. Cs\'aki, J. Erlich, G. D. Kribs, and J. Terning,
Phys. Rev. D \textbf{66}, 075008 (2002).
\bibitem{extrad} S. Dimopoulos, D. E. Kaplan, and N. Weiner,
Phys. Lett. \textbf{B534}, 124 (2002);
T. Li and W. Liao, Phys. Rev. \textbf{B545}, 147(2002);
\bibitem{orbifoldsu3} W. Chang and J. N. Ng, Phys. Rev. D \textbf{69}, 056005
(2004). 
 \bibitem{orbifold331} I. Gogoladze, Y. Mimura, and S. Nandi,
Phys. Lett. \textbf{B554}, 81 (2003).
 
\bibitem{outros331} R. Martinez, W. A. Ponce, and
L. A. S\'anchez, Phys. Rev. D \textbf{65}, 055013 (2002); W. A. Ponce,
Y. Giraldo, and L. A. S\'anchez, Phys. Rev. D \textbf{67} (2003) 075001.


\bibitem{assi} J. C. Montero, V. Pleitez, and M. C. Rodriguez,
Phys. Rev. D \textbf{58}, 094076, 097505 (1998),  Int. J. Mod. Phys.
\textbf{A16}, 
1147, 2001; and hep-ph/0204130; P. H. Frampton and A. Rasin, Phys. Lett. {\bf
B482}, 129 (2000).
\bibitem{tavares} M. A. Perez, G. Tavares-Velasco and J. J. Toscano, Phys. Rev. 
D \textbf{69}, 115004 (2004).
\bibitem{willmann99} L. Willmann {it et. al.}, Phys. Rev. Lett. \textbf{82}, 49
(1999).
\bibitem{pleitez00} V. Pleitez, Phys. Rev. D \textbf{61}, 057903 (2000).
\bibitem{dion} B. Dion, T. Gregoire, D. London, L. Marleau and H. Nadeau,
Phys. Rev. D \textbf{59}, 075006 (1999).
\bibitem{maisbi} M. B. Tully and G. C. Joshi, Phys. Lett. \textbf{B466}, 333
(1999); E. M. Gregores, A. Gusso and S. F. Novaes, Phys. Rev. D
\textbf{64}, 015004 (2001);
J. L. Garcia-Luna, G. Tavares-Velasco, J. J. Toscano, Phys. Rev. D \textbf{69},
093005 (2004).
%%polo de landau a trivialidade

\bibitem{landau} L. D. Landau, in \textsl{Niels Bohr and the Development of
Physics}, edited by W. Pauli (Pergamon Press, London, 1955); p. 52, and
references therein.
\bibitem{qed1} M. G\"ockerler, \textit{et al.}, Phys. Rev. Lett. \textbf{19},
4119 (1998).
\bibitem{qed2} H. Gies and J. Jaeckel, hep-ph/0405183.

\bibitem{swbook} S. Weinberg, \textsl{The Quantum Theory of Fields}, (Cambridge
University Press, New York, 1996); Sec. 18.3.
\bibitem{hambye} R. Dashen and H. Neuberger, Phys. Rev. Lett. \textbf{24}, 1897 
(1983); M. A. B. B\'eg, C. Panagiotakopoulos and A. Sirlin, Phys. Rev. Lett.
\textbf{52}, 883 (1984); M. Lindner, Z. Phys. \textbf{C31}, 295 (1986); 
J. Kuti, L. Lin and Y. Shen, Phys. Rev. Lett. \textbf{61}, 678 (1988);
P. Hasenfratz and J. Nager, Z. Phys. \textbf{C 37}, 477 (1988); M. L\"uscher and
P. Weisz, Phys. Lett. \textbf{B212}, 472 (1988); T. Hambye and K. Riesselmann,
Phys. Rev. D, \textbf{55}, 7255 (1997), and references therein.
\bibitem{holland} K. Holland, hep-lat/0409112.
\bibitem{ng94} D. Ng, Phys. Rev. D \textbf{49}, 4805 (1994).
\bibitem{jain94} P. Jain and S. D. Joglekar, Phys. Lett. \textbf{B407}, 151 
(1997). 
\bibitem{phf98} P. H. Frampton, Int. J. Mod. Phys. \textbf{A13}, 2345
(1998). More recently a value of 4 TeV for the energy of the 
pole has been given in P. H. Frampton, Mod. Phys. Lett. \textbf{A18}, 1377
(2003).

\bibitem{cp3} J. C. Montero, V. Pleitez and O. Ravinez, Phys. Rev. D
\textbf{60}, 076003 (1999).
\bibitem{pdg}  S. Eidelman {\textit et al.} (Particle Data Group), Phys. Lett.
{\bf B592}, 1 (2004). 

\bibitem{jones} D. R. T. Jones, Phys. Rev. D \textbf{25}, 581 (1982); M. E.
Machacek and M. T. Vaugh, Nucl. Phys. \textbf{B222}, 83 (1983).
\bibitem{dasjain} P. Das and P. Jain, Phys. Rev. D \textbf{69}, 075001 (2000).
\bibitem{mpp} J. C. Montero, F. Pisano, and V. Pleitez, Phys. Rev. D
\textbf{47}, 2918 (1993); R. Foot, H. N. Long, and T. A. Tran, Phys. Rev. D
\textbf{50}, R34 (1994); H. N. Long, \textit{ibid}. D \textbf{54}, 4691 (1996).
\bibitem{tavares2} G. Tavares-Velasco and J. J. Toscano, Phys. Rev.
D\textbf{70}, 053006 (2004), hep-ph/0407047. 
\bibitem{su4} F. Pisano and V. Pleitez, Phys. Rev. D \textbf{51}, 3865 (1995).

\end{thebibliography}
\end{document}